# PHOTON SPLITTING IN STRONGLY MAGNETIZED COSMIC OBJECTS – REVISITED

Günter Wunner, Robert Sang and Dagmar Berg

Theoretische Physik I, Ruhr-Universität Bochum, D-44780 Bochum, Germany

## ABSTRACT

We reassess the importance of photon splitting to spectral formation in cosmic X-ray and $\gamma$-ray sources with magnetic fields of several $10^{12}$ gauss. Our analysis is based on the recent numerically accurate evaluation of the full relativistic expression for photon splitting at neutron star magnetic field strengths (Mentzel et al. 1994). Using these results we demonstrate that Adler's (1971) widely used approximation formula for photon splitting underestimates the correct splitting rate by *several* orders of magnitude, at these magnetic field strengths. Even down at X-ray energies of a few tens of keV magnetic photon splitting is found to be an efficient photon conversion process, producing mean free paths as short as several tens of centimeters for fields on the order of $5 \times 10^{12}$ gauss. We conclude that previous analyses of the contributions of magnetic photon splitting to spectral formation based on Adler's approximation formula are physically unreliable, and, in particular, that photon splitting must be included in realistic modelings of spectra of both strongly magnetized cosmic X-ray and $\gamma$-ray sources.

## 1. INTRODUCTION

It has always been one of the most intriguing features of the superstrong magnetic fields on the order of $10^{12}$ to $10^{13}$ gauss discovered in the vicinity of neutron stars that elementary physical processes that are forbidden in free space become allowed. A famous example is the production of an electron-positron pair by a *single* high energy photon in the presence of a strong field. This process is believed to be a central mechanism underlying the generation of high energy radiation in compact strongly magnetized astrophysical sources. A review of the importance of this process and other exotic processes to high energy astrophysics can be found in the book by Mészáros (1992). The reason why new processes become allowed is that, loosely speaking, the strong magnetic field is able to "absorb" momentum perpendicular to the field, with the result that the law of conservation of momentum needs to be fulfilled only in one spatial dimension, viz. parallel to the field. More precisely, the motion of electrons is no longer described by plane waves in three dimensions but by the quantized Landau state solutions of Dirac's equation in a magnetic field.

In this *Letter* we want to turn the attention to another exotic process, namely the decay of a single high energy photon into two lower energy photons in the presence of a strong magnetic field (magnetic photon splitting). It has been pointed out (Harding 1991) that below the pair creation threshold of $2\,m_e = 1.022$ MeV this would be the dominant process which limits the lifetime of a high energy photon, and Baring suggested the importance of photon splitting as a possible cooling mechanism for hot photons in $\gamma$-ray burst sources (1991) and soft $\gamma$-repeaters (1995). However, all previous studies of the relevance of photon splitting to the physics of strongly magnetized cosmic high energy sources suffered from the fact that no quantitative results existed for the very complicated exact cross section of photon splitting, and therefore discussions of the astrophysical implications of magnetic photon splitting had to rely on simple analytical expressions derived by Adler (1971) and Papanyan and Ritus (1972) valid only in the weak-field limit $B \ll B_{\rm cr}$,

$$R(\omega) \approx 0.12 \, \left(\frac{B}{B_{\rm cr}}\right)^6 \, \left(\frac{\omega}{m_e}\right)^5 \, {\rm cm}^{-1} \,. \tag{1}$$

Here $R(\omega)$ is the attenuation coefficient, i.e. the inverse of the mean free path, $B_{\rm cr} = m_e^2/e = 4.414 \times 10^{13}$ gauss is the reference field strength, where the cyclotron energy becomes equal to the rest energy of an electron, and for convenience we have set $\hbar = c = 1$.

This situation has changed recently in that a numerical evaluation of the exact cross section of photon splitting at realistic neutron star magnetic field values has become available (Mentzel et al. 1994). It is the purpose of this *Letter* to reassess the importance of magnetic photon splitting to high energy astrophysics on the basis of these new results. We do so by comparing magnetic photon splitting below the pair creation threshold with single photon pair production above the threshold. We find that the two processes have mean free paths of comparable sizes at neutron star magnetic field strengths of several $10^{12}$ gauss, and that the simple analytical expressions used previously underestimate the efficiency of photon splitting at these field strengths by several orders of magnitude. As a consequence, previous investigations of the contributions of magnetic photon splitting to spectral formation in strongly magnetized $\gamma$-sources must be repeated using the new accurate numerical data for photon splitting, and more generally, photon splitting must be consistently implemented in modelings of the spectra of both soft $\gamma$-ray and hard X-ray sources.

## 2. RESULTS AND DISCUSSION

Without loss of generality we will restrict our considerations to photons propagating perpendicular to the direction of the magnetic field ($k_z = 0$). It immediately follows from the behavior of photons under Lorentz transformations that the attenuation coefficient $R'$ of a photon propagating at an arbitrary angle $\theta'$ with respect to the magnetic field axis and with frequency $\omega'$ is related to the attenuation coefficient $R(\omega, \pi/2)$ in the $k_z = 0$ frame by

$$R'(\omega', \theta') = \sin\theta' \ R(\omega' \sin\theta', \pi/2). \qquad (2)$$

The quantum electrodynamical derivations of the cross sections of single photon pair production and photon splitting in magnetic fields have been treated in great detail in the literature and there is no need to repeat the complicated formulae for the cross sections here (for photon splitting see e.g. Adler 1971, Stoneham 1979, Mentzel et al. 1994, for single photon pair production see e.g. Daugherty and Harding 1983 and references therein, for reviews see Harding 1991 and Mészáros 1992). Therefore we will immediately turn to the discussion of the results of the numerical evaluations of these cross sections and their implications for the modeling of strongly magnetized astrophysical sources.

In Fig. 1 we confront, for the quantum electrodynamical reference field strength $B_{\rm cr}$, the attenuation coefficient for photon splitting *below* the pair creation threshold with the attenuation coefficient for $1\gamma$ pair creation *above* the threshold. The coefficient for photon splitting has been obtained by a numerical evaluation of the exact relativistic expression for the photon splitting rate as derived by Mentzel et al. (1994). The results for $1\gamma$ pair creation above the threshold were computed by evaluating the expression given by Daugherty and Harding (1983) (Eq. (5) of their paper, cf. also their Fig. 2). The sawtooth-like behavior of the $1\gamma$ pair creation cross section evident in Fig. 1 is well known from the literature: spikes occur whenever a new Landau channel opens and an electron-positron pair can be created "at rest", i.e. with zero momentum along the field direction. (The singularities are caused by the one-dimensional density of states of the electrons and positrons in Landau states, and behave as $1/\sqrt{\varepsilon}$, where $\varepsilon$ is the energy distance from the the respective threshold). In Fig. 1 we have averaged over the polarization of the incoming photon, but the conclusions to be drawn below are valid also for polarized photons.

We see in Fig. 1 that the attenuation coefficient for photon splitting exhibits a sharp decrease for $\omega \to 0$, shows an almost plateau-like behavior in the intermediate energy range, and diverges at the pair creation threshold (the corresponding $S$-matrix element diverges as $1/\sqrt{(2m_{\rm e} - \omega)}$, cf. Mentzel et al. 1994). Comparing the average size of the attenuation coefficient for photon splitting in the range between 0 and $2\,m_{\rm e}$ with the size of the coefficient for pair creation above the threshold we recognize that the first is smaller than the latter by roughly four orders of magnitude. This is expected since photon splitting is a third-order process, and therefore it should be less probable than the first-order process of $1\gamma$ pair creation by the second power of the fine structure constant ($\alpha^2 \sim (1/137)^2 \sim \frac{1}{2} \times 10^{-4}$).

The surprise comes when one looks at the corresponding results for magnetic field strengths smaller by one order of magnitude, $B \sim 0.1\,B_{\rm cr}$, i.e. field strengths that have actually been observed in radio and X-ray pulsars and, possibly, in $\gamma$-ray burst sources. Figure 2 shows the results for $B = 0.1\,B_{\rm cr}$. While every cross section for itself qualitatively possesses the same overall behavior as for $B_{\rm cr}$, we find that now the average sizes of the attenuation coefficients for photon splitting below the threshold and for pair creation above it are approximately of the same order of magnitude.

This behavior can be understood as follows. The cross section for pair creation contains a general pre-factor of the form $\exp(-1/B)$, expressing the fact that this process becomes kinematically forbidden as $B$ goes to 0 (the exponential dependence on $B$ explains the very much reduced absolute size of the attenuation coefficient for pair production in Fig. 2 as compared to Fig. 1). The $S$-matrix elements for photon splitting, on the other hand, are coherent sums over many intermediate states involving products of matrix elements each of which contain exponential $B$ dependences, but obviously at these magnetic field strengths the many summations are capable of compensating for the exponential dependence of the individual terms. This produces a magnetic field dependence that is weaker than exponential. In fact the weak-field, low-frequency limit of magnetic photon splitting rate, Eq. (1), shows that as $B \to 0$ the rate goes to zero with a power of $B$, and not exponentially.

The predictions of the approximation formula (1) are rendered in Figures 1 and 2 by the dotted curves, and can be compared with the accurate results. It can be seen that even though the frequency dependence is qualitatively correctly reproduced over the range 0 to $2\,m_{\rm e}$ the absolute sizes of the rates are underestimated by several orders of magnitude, at the photon energies resolved in the figures. Evidently at these field strengths the range of applicability of the weak-field, low-frequency form of the exact expression for photon splitting is restricted to much smaller photon energies than was previously assumed. (A possible explanation is that the expansion of the exact rate in terms of the photon energy and the field strength about the nonanalytic point $B = 0$ is an asymptotic one, in which taking into account only the leading term may not be sufficient.) Consequently, analyses of the contributions of magnetic photon splitting to spectral formation in compact magnetic cosmic objects which used the approximation formula are necessarily unreliable, and should be repeated using the correct rates.

## 3. CONCLUSIONS

The important physical conclusion to be drawn from our analysis for the physics of neutron stars with surface fields on the order of several $10^{12}$ gauss is that photon splitting, as a lifetime limiting process of $\gamma$ quanta, is effective below $2\,m_e$ to the same extent as is single photon pair creation above $2\,m_e$. It can be read off Fig. 2 ($B = 0.1\,B_{cr}$) that typical mean free paths for photon splitting below 1 MeV are of the order of millimeters, but even down at X-ray energies of 5 to 50 keV photon splitting is so efficient as to produce mean free paths on the order of tens of centimeters. This is clear indication that photon splitting at neutron star magnetic field strengths is not just an exotic, negligible third-order process, but should play an important role in the reprocessing of photons with energies below $2\,m_e$ in compact astrophysical objects with strong magnetic fields. In particular, it should be taken into account in realistic modelings of the spectra of magnetic X-ray and $\gamma$-ray sources.

We thank Markus Mentzel for many helpful discussions and comments.

---

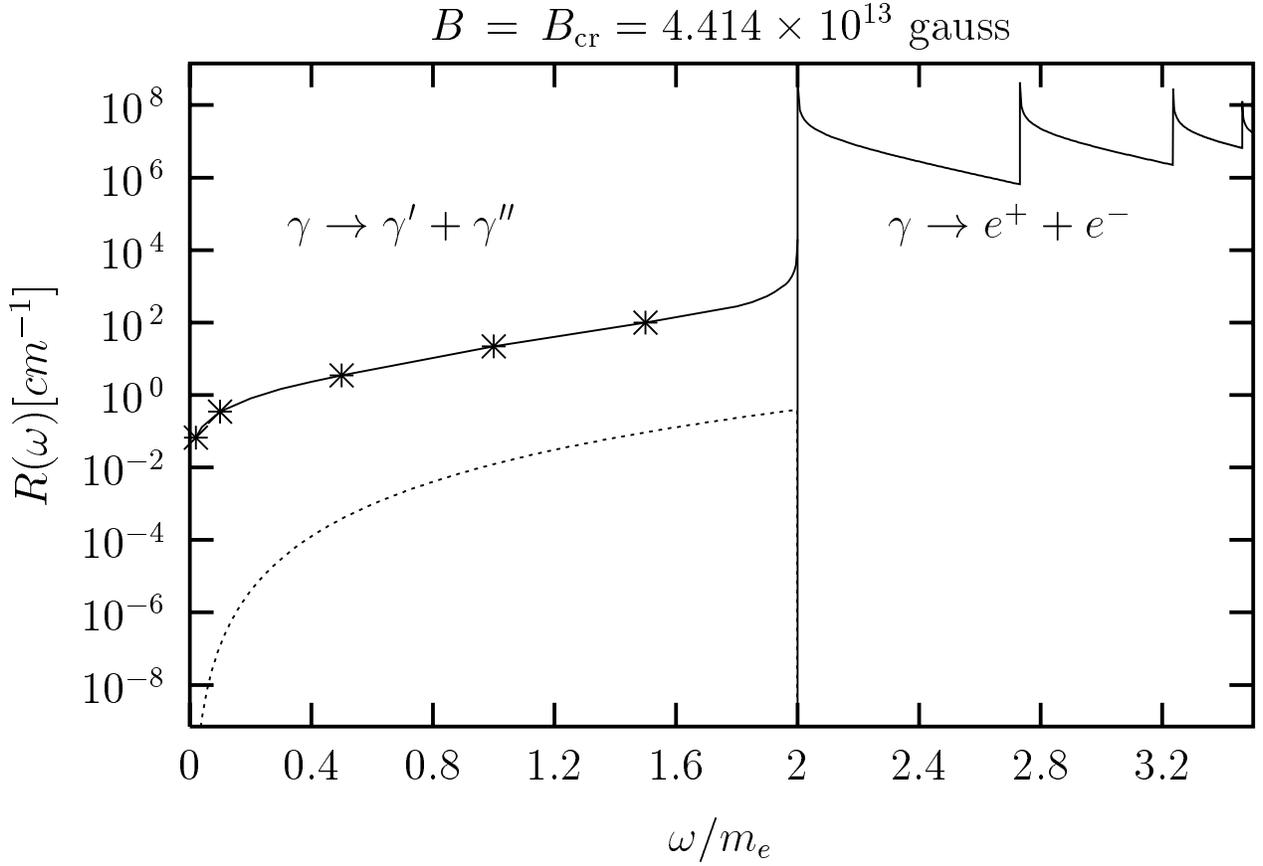

Fig. 1.— For a magnetic field strength of $B = B_{\rm cr} = 4.414 \times 10^{13}$ gauss, the figure shows a comparison between the exact attenuation coefficient $R(\omega)$ (inverse mean free path) for photon splitting ($\gamma \to \gamma' + \gamma''$) below the pair production threshold, $2\,m_{\rm e} = 1.022$ MeV, with the attenuation coefficient for pair production ($\gamma \to e^+ + e^-$) above the threshold, for unpolarized photons propagating perpendicular to the magnetic field. The results have been obtained evaluating the full relativistic expressions given by Mentzel et al. (1994) for photon splitting, and by Daugherty and Harding (1983) for pair production. Also shown is the prediction of Adler's (1971) approximation formula for photon splitting (dotted curve). Obviously the latter underestimates the accurate results by several orders of magnitude.

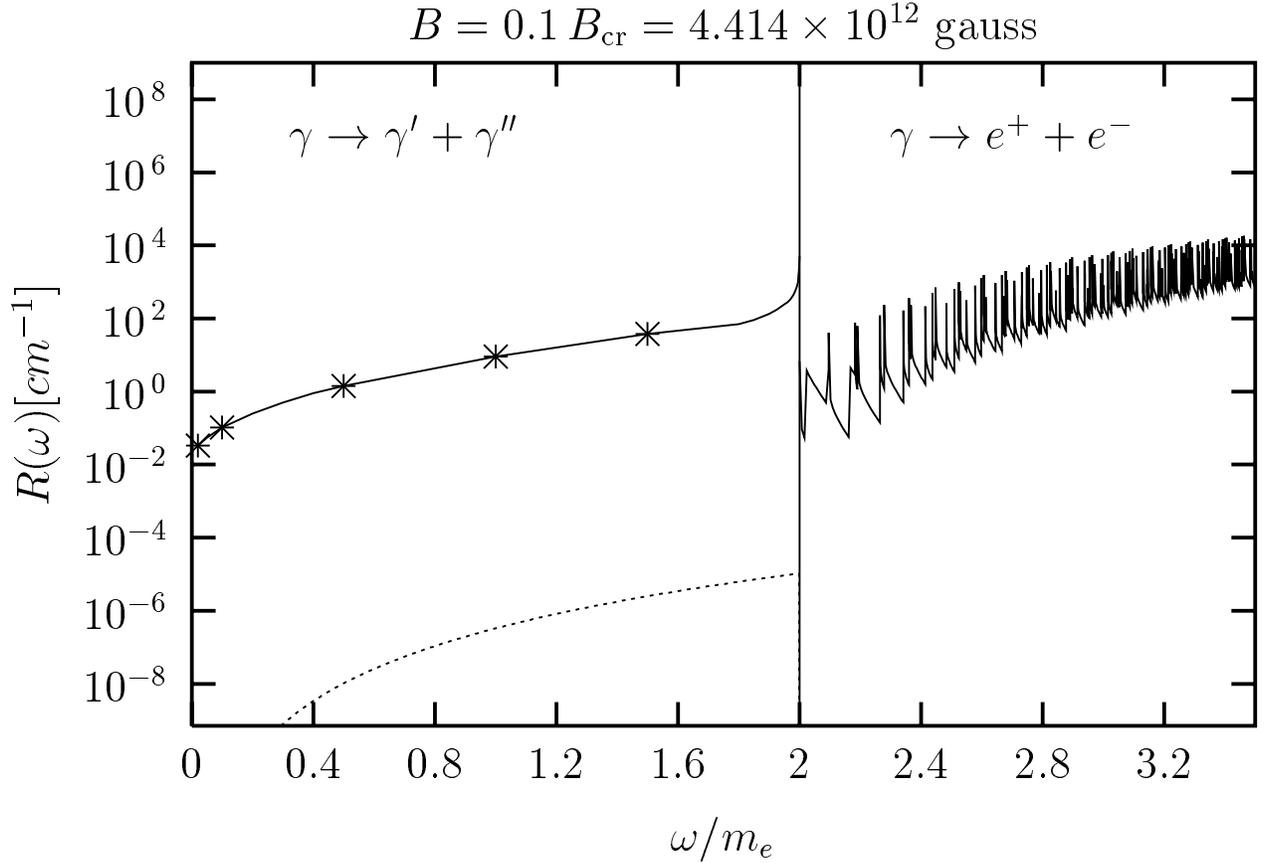

Fig. 2.— Same as Fig. 1, but for a magnetic field strength of $B = 0.1\,B_{\mathrm{cr}} = 4.414 \times 10^{12}$ gauss. Note that at this field strength the mean free paths for photon splitting below the threshold and for pair production above the threshold are approximately of the same order of magnitude. Again, the low-field, low-frequency approximation formula for photon splitting (dotted curve) deviates from the accurate curve by several orders of magnitude, at finite values of $\omega$.